\documentstyle{mn}
\onecolumn
\input{epsf}
\newif\ifAMStwofonts

\def\ltsima{$\; \buildrel < \over \sim \;$}
\def\simlt{\lower.5ex\hbox{\ltsima}}
\def\gtsima{$\; \buildrel > \over \sim \;$}
\def\simgt{\lower.5ex\hbox{\gtsima}}

\title[Statistics of Collapsed Objects: II]
{From Snakes to Stars, the Statistics of Collapsed Objects - II.
Testing a Generic Scaling {\em Ansatz} for Hierarchical
Clustering}

\author[D. Munshi et al.]{Dipak Munshi$^{1,4}$,
Peter Coles$^{1,3}$ and Adrian L. Melott$^2$,\\ $^1$Queen Mary and
Westfield College, London E1 4NS, United Kingdom \\ $^2$Department
of Physics and Astronomy, University of Kansas, Lawrence, Kansas
66045, U.S.A., \\ $^3$School of Physics and Astronomy, University
of Nottingham, University Park, Nottingham NG7 2RD, United
Kingdom\\ $^4$International School for Advanced Studies (SISSA),
Via Beirut 2-4, I-34013 Trieste, Italy}
\pagerange{000,000}
\pubyear{1996}

\begin{document}

\maketitle

\begin{abstract}
We develop a diagrammatic technique to represent the multi-point
cumulative probability density function (CPDF) of mass
fluctuations in terms of the statistical properties  of individual
collapsed objects and relate this to other statistical descriptors
such as cumulants, cumulant correlators and factorial moments. We
use this approach to establish key scaling relations describing
various measurable statistical quantities if clustering follows a
simple general scaling {\em ansatz}, as expected in hierarchical
models. We test these detailed predictions against high-resolution
numerical simulations. We show that, when appropriate variables
are used, the count probability distribution function (CPDF) shows
clear scaling properties in the non-linear regime. We also show
that analytic predictions made using the scaling model for the
behaviour of the void probability function (VPF) also match the
simulations very well.  We generalise the results for the CPDF to
the two-point (bivariate) count probability distribution function
(2CPDF), and show that its behaviour in the simulations is also
well described by the theoretical model, as is the bivariate void
probability function (2VPF). We explore the behaviour of the  bias
associated with collapsed objects in limit of large separations,
finding that it depends only on the intrinsic scaling parameter
associated with collapsed objects and that the bias for two
different objects can be expressed as a product of the individual
biases of the objects. Having thus established the validity of the
scaling ansatz in various different contexts, we use its
consequences to develop a novel technique for correcting
finite-volume effects in the estimation of multi-point statistical
quantities from observational data.
\end{abstract}


\section{Introduction}
The process of gravitational instability leads to the growth and
collapse of initially small inhomogeneities in a self-gravitating
medium. This is the basic idea behind theories for the formation
of galaxies and large-scale structures in the Universe, as the
small initial fluctuations seed the formation of collapsed objects
that progressively cluster and merge into objects of larger size,
such as galaxies and clusters. Much progress has been made during
the last two decades using numerical simulations to follow the
evolution of cosmological density fluctuations up to and beyond
their collapse into bound structures, but it is nevertheless
important to develop an analytical framework for describing this
phenomenon and understanding what the simulation results.

Much of the previous analytic work in this direction has been
based on an extended version of the Press-Schechter (1974) model.
Such methods have been used to study the distribution of formation
epochs (Bond et al. 1991; Lacey \& Cole 1993; see White 1993 for a
complete review), merger rates and survival times of haloes and
also how the distribution of collapsed objects  relates to the
underlying mass distribution (Mo \& White 1996; Mo et al. 1997).
Tests of these predictions have shown good agreement with
theoretical model predictions. Other recent methods which try to
explain statistics of collapsed dark objects depend on extension
of Zel'dovich approximation (Lee \& Shandarin 1997, 1998; Catelan
et al. 1998; Porciani et al. 1998).

A parallel approach has been to assume that statistical properties
of the distribution of collapsed objects - particularly the
many-body correlation functions -  follow a hierarchical scaling
pattern (Balian \& Schaeffer 1989a,b). This kind of scaling can be
used to predict, for example,  the mass function of collapsed
objects (Valageas \& Schaeffer 1997) and their correlation
properties (Bernardeau \& Schaeffer 1992; Munshi et al. 1998b).
These results are especially interesting because they relate
directly to the statistics of the underlying mass distribution and
hence to the dynamical origin of the distribution governed by the
BBGKY equations.

Predictions of the behaviour of one-point statistical quantities
such as the count probability distribution function (CPDF) or void
probability distribution function (VPF) for the underlying mass
distribution in the highly non-linear regime of strong clustering
were derived using this kind of scaling {\em ansatz} by Balian \&
Schaeffer (1988, 1989). These results  were later extended to the
quasi-linear regime where a perturbative series expansion can be
used (Beranardeau 1992, 1994a, 1994b, 1995). It emerged from these
and other studies that a scaling hierarchy naturally developed by
the action of gravitational clustering on Gaussian (random-phase)
initial perturbations. Predictions for these one-point quantities
have been tested with reasonable success against observations
using various galaxy catalogs (Maurogordato \& Lachieze-Rey 1987;
Blanchard et al. 1990; Mourogordato et al. 1992;  Bouchet et al.
1993) and numerical simulations (Bouchet \& Hernquist 1992;
Lucchin et. al 1994;
Colombi et al. 1992, 1994, 1995, 1996; Munshi et al. 1998c),
although Lucchin et al. (1994) and Colombi et al. (1996) found some
deviation from scaling.

It is natural now to ask whether these properties of one-point
statistical quantities can be generalised to two-point statistics,
or even beyond. The two-point count probability distribution
function (2CPDF) and two-point void probability function (2VPF),
both defined in detail below, are particularly interesting for a
number of different reasons. The first is that the 2VPF can be
shown to be a generating function for the cumulant correlators
(CCs) in a way which is very similar to the VPF, which acts as a
generating function for cumulants. Recently it was shown that CCs
can be reliably measured using galaxy catalogs (Szapudi et al.
1992; Meiksin et al. 1992; Szapudi et al 1995; Szapudi \& Szalay
1997). They were also measured in numerical simulations (Munshi \&
Melott 1998) to test extended perturbation theory (EPT) as
proposed by Colombi et al. (1996). Since it becomes increasingly
difficult to measure these quantities with increasing order, it is
sensible instead to study their generating function (2VPF) which
carries information to all orders. The second reason is that these
two-point objects are useful in the problem of estimating errors
associated with the determination of one-point quantities such as
the CPDF, VPF and $S_N$ parameters (Szapudi \& Colombi 1995).
Although analytical results related to two-point quantities are
frequently used to estimate and correct such errors, no systematic
studies have yet been done to test these predictions directly
against numerical simulations, partly because of the difficulty of
performing numerical simulations with the very large dynamic range
necessary to study these quantities. Finally, the 2CPDF carries
information as to how overdense  regions are biased tracers of the
underlying mass distribution, so a computation of the bias implied
by the scaling ansatz can provide an understanding of the origin
of bias that can be compared with results from other approaches.

Before proceeding, however, we stress that one should not get too
carried away by the success of this kind of scaling model. The
hierarchical ansatz is, at best, a simplifying hypothesis which,
admittedly  seems to work very well for ``realistic models'' of
structure formation. But no firm connection has yet been
established with the  microscopic physics of collisionless
clustering described by BBGKY equations (Peebles 1980).
Most efforts in solving the BBGKY equations have
focused on particular closure schemes (Davis \& Peebles 1977), the
general separability of position correlations and momentum
correlation in phase-space (Fry 1984;  Hamilton 1988) and its
stability properties (Ruamsuwan \& Fry 1992; Yano \& Gouda 1998).
Although these precious efforts in solving the highly complicated
nonlinear integro-differential BBGKY equations provide us with
valuable insights, they also help underline the complications in
arriving in any general solution to the problem. Our inability to
solve these equations in the highly non-linear regime leaves us
with undetermined parameters which can only be computed from
numerical simulations, thus complicating the task of testing the
scaling ansatz itself.

This paper is structured in the following way. Section 2 is
devoted to a theoretical discussion of multi-point cumulants,
multi-point factorial moments, multi-point factorial correlators
and their various generating functions.   We also develop a
general approach to the treatment of finite-volume effects using
these quantities. Section 3 is devoted specifically to one-point
and two-point statistics. We summarize the main analytical results
of Balian \& Schaeffer (1989) and Bernardeau \& Schaeffer (1992)
and test them numerically. Section 4 contains a detailed
description of our numerical simulations and to details of data
analysis techniques; we also present the results of our analysis
in this section. We discuss the main results and relate it to
other studies in Section 5.

\section{Voids \& Probabilities from Cumulant Correlators}

The void probability function (VPF) is the probability that a
randomly-placed sphere of some volume $V$ contains no galaxies.
This can be written
\begin{equation}
 P_V(0) = \exp \left(-{\phi(N_c) \over \bar \xi_2}\right),
 \end{equation}
 in which the mean number of galaxies per cell is $\bar{N}$
 and the volume-average of the two-point correlation
function over the cell is
\begin{equation}
\bar{\xi}_2=\frac{1}{V^2}\int \int \xi_2({\bf r}_1, {\bf r}_2)
dV_1 dV_2.
\end{equation}
The quantity $N_c=\bar{N}\bar{\xi}_2$.  The VPF is a generating
function of $P(N)$, the count probability distribution function
(CPDF) for cells of the same size (Balian \& Schaeffer 1989). Let
us assume  Using the generating function relationship between the
CPDF and VPF we can write
\begin{equation}
P(N) = {(-1)^N \over N!} {d \over d\mu^N} \exp \left( - {\phi(\mu
N_c) \over \bar \xi_2} \right)\bigg |_{\mu = 1},
\end{equation}
which can also be written in the following integral form (Munshi
et al. 1998c)
\begin{equation}
P(N) = {1 \over 2\pi i}  \int { d \lambda \over \lambda^{N+1}}
\exp \left( - { \phi ((1 - \lambda) N_c) \over \bar \xi_2} \right).
\end{equation}
The integral is to be evaluated along a contour around $\lambda =
0$. We also  define  the function $\Psi(t) = - \phi(-t)$ which can
be directly related to the function $\Pi(\nu)$, the continuous
analogue of $P(N)$,
\begin{equation}
\exp \left({ \Psi(t) \over \bar \xi_2 }\right) = \int_0^{\infty} d
\nu \exp \left( {\nu t \over N_c} \right) \Pi( \nu).
\end{equation}
Using the definition of $P(N)$ now one can easily show that
\begin{equation}
P(N) = \int_0^{\infty} d \nu { \exp (-\nu) \nu^N \over N! } \Pi( \nu).
\end{equation}
This is the form of distribution obtained when $\Pi(\nu)$ is
Poisson sampled, because it is the convolution of the function
$\Pi(\nu)$ with a Poisson distribution describing the shot noise
effects. This demonstrates that $\Pi(\nu)$ can be viewed as the
continuous limit of $P(N)$ in the limit of large number densities.
This can be seen by change of variable $\lambda = 1 + t/N_c$ in
equation (3) and then taking the limit $N_c \rightarrow \infty$
and $N \rightarrow \infty$ with the ratio $N/N_c$ remaining
finite; this also gives $ P(N) = \Pi(N)$.

The factorial moments of $P(N)$ can now be related to the moments
of $\Pi(\nu)$ with the aid of the definition of the $p$-th
factorial moment:
\begin{equation}
\langle (N_1)_p \rangle = \sum_{N=0}^{\infty} N(N-1)\dots(N-p+1)P(N)= \int_0^{\infty} \nu^p
\Pi(\nu) d\nu.
\end{equation}
One can similarly define the normalized factorial moments of
$P(N)$ by
\begin{equation}
\Sigma_p = {\bar \xi_2 \over p! N_c^p} \sum N(N-1)... (N-p+1)P(N) =
{\bar \xi_2 \over p! N_c^p} (N_1)_p.
\end{equation}
One can define a generating function $\Sigma(t)$ for the
$\Sigma_p$ parameters using
\begin{equation}
 \Sigma(t) = \sum_1^{\infty} \Sigma_p t^p
 \end{equation}
 and we have the following expression connecting
these two generating functions $\Psi(t)$ and $\Sigma(t)$ (Munshi
et al. 1998c),
\begin{equation}
1 + {\Sigma(t) \over \bar \xi_2} = \exp ( {\Psi(t) \over \bar \xi_2} ),
\end{equation}
which can also be written as
\begin{equation}
\Psi(t) = \bar \xi_2 \ln ( 1 + {\Sigma(t) \over \bar \xi_2} ).
\end{equation}

It is possible to extend the method of generating functions to
multi-point factorial moments. If we denote the two-point CPDF
$P(N_1,N_2)$, the joint probability of finding $N_1$ particles in
the  first cell and $N_2$ particles in a second cell, and assume
locally Poisson sampling distribution then we can relate
$P(N_1,N_2)$ with its smooth counterpart $\pi(\nu_1,\nu_2)$ by:
\begin{equation}
P(N_1, N_2) = \int_0^{\infty} d\nu_1 \int_0^{\infty} d\nu_2
{ \exp(-\nu_1) \nu_1^{N_1} \over N_1! } { \exp(-\nu_2) \nu_2^{N_2} \over N_2! }
\Pi (\nu_1, \nu_2)
\end{equation}
so that the two-point factorial moment becomes
\begin{equation}
\langle (N_1N_2)_{pq} \rangle= \sum_{N=0}^{\infty} N_1(N_1-1)\dots(N_1-p+1)N_2(N_2-1)\dots(N_2-q+1)P(N_1,N_2)=
\int_0^{\infty} \nu_1^p d \nu_1  \int_0^{\infty}\nu_2^q   d\nu_2  ~\Pi (\nu_1, \nu_2)
\end{equation}
We define the normalised two-point factorial moment by following
equation.
\begin{equation}
\Sigma^{(2)}_{pq} = {{\bar \xi_2}^2 \over p! q! N_c^{p+q}} \sum N_1(N_1-1)
\dots(N_1-p+1)N_2(N_2-1)\dots(N_2-q+1)P(N_1, N_2) = {{\bar \xi_2}^2 \over p! q! N_c^{p+q}}
(N_1N_2)_{pq}
\end{equation}
It is to be noted that, in general, owing to statistical isotropy
and homogeneity $\Sigma_{pq}$ should depend only on the separation
of two cells. We define the generating function for $\Sigma_{pq}$
by $\Sigma^{(2)}$, i.e.
\begin{equation}
\Sigma^{(2)} = \sum_{p,q=1}^{\infty}\Sigma_{pq} t_1^p t_2^q
\end{equation}
and similarly the generating function for the 2CCs by
\begin{equation}
{\Psi^{(2)}(t_1, t_2) \over \bar \xi_2^2} = \sum {t_1^p t_2^q
\over p! q!} {\langle \delta^p(x_1)\delta^q(x_2) \rangle \over
\langle \delta^2(x) \rangle^{(p+q)}}.
\end{equation}
Using similar technique as used for the case of one-point moments
we can relate $\Sigma^{(2)}(t_1, t_2)$ with $\Psi^{(2)}(t_1,t_2)$:
\begin{equation}
1 + {\Sigma(t_1) \over \bar \xi_2} + {\Sigma(t_2) \over \bar \xi_2} +
{\Sigma^{(2)}(t_1,t_2) \over {\bar \xi_2}^2} = \exp \left( { \Psi(t_1)
\over \bar \xi_2} +  { \Psi(t_2) \over \bar \xi_2} + {\Psi^{(2)}(t_1, t_2)
\over {\bar \xi_2}^2} \right)
\end{equation}
A Taylor expansion of the above equation provides us with a set of
relations which can be used to measure $\Psi^{(2)}_{pq}$ once
$\Sigma^{(2)}_{pq}$ is computed:
\begin{eqnarray}
\Psi^{(2)}_{11} & = & (-1 + \Sigma^{(2)}_{11}) \\ \Psi^{(2)}_{12}
& =  & {1 \over {\bar \xi_2}}(1 - \Sigma^{(2)}_{11} + {\bar
\xi}_2(\Sigma^{(2)}_{12} - \Sigma_{2})
\\ \Psi^{(2)}_{22} & = &{1 \over {\bar \xi_2}^2}( -3 +
4\Sigma^{(2)}_{11} - {\Sigma^{(2)}_{11}}^2 -
2\xi_2(\Sigma^{(2)}_{12} - 2\Sigma_{2} + \Sigma^{(2)}_{21}) - 2
\xi_2^2( \Sigma_{2}^2 - \Sigma^{(2)}_{22}))
\\ \Psi^{(2)}_{13} & = & {1 \over {\bar \xi_2}^2}( -1 +
\Sigma^{(2)}_{11} - {\bar \xi}_2(\Sigma^{(2)}_{12} - 2\Sigma_{2})
- {\bar \xi}_2 \Sigma^{(2)}_{11} \Sigma_{2} + {\bar
\xi}_2^2(\Sigma^{(2)}_{13} - \Sigma_{3}))\\ \Psi^{(2)}_{23} & = &
{1 \over {\bar \xi}_2^3}( 2 + {\Sigma^{(2)}_{11}}^2 - \Sigma_{11}(
3 + {\bar \xi}_2( \Sigma^{(2)}_{12} - 2 \Sigma_{2})) \\
\nonumber&& + {\bar \xi}_2(2 \Sigma^{(2)}_{12} - 4 \Sigma_{2} +
\Sigma^{(2)}_{21})) - {\bar \xi}_2^2(\Sigma^{(2)}_{13} -
2\Sigma_{2}^2 + \Sigma_{2}\Sigma^{(2)}_{21} + \Sigma_{22} -
\Sigma_3) + {\bar \xi}_2^3(\Sigma_{23} - \Sigma_2\Sigma_3))
\end{eqnarray}
In our above analysis we have used two different cells of the same
size, but it is now obvious that the corresponding result for two
cells of different size will read
\begin{equation}
1 + {\Sigma(t_1) \over \bar \xi_2^x} + {\Sigma(t_2) \over \bar \xi_2^y} +
{\Sigma^{(2)}(t_1,t_2) \over {\bar \xi_2^x} {\bar \xi_2^y}} = \exp \left(
{\Psi(t_1) \over \bar \xi_2^x} +  { \Psi(t_2) \over \bar \xi_2^y} +
{\Psi^{(2)}(t_1, t_2) \over {\bar \xi_2^x} {\bar \xi_2^y}} \right).
\end{equation}
These results can be extended reasonably straightforwardly to
three-point quantities:
\begin{equation}
1 +{\Sigma(t_1) \over \bar \xi_2} + {\Sigma(t_2) \over \bar \xi_2}
+ {\Sigma(t_3) \over \bar \xi_2} + {\Sigma^{(2)}(t_1,t_2) \over
\bar \xi_2^2} + {\Sigma^{(2)}(t_1,t_3) \over \bar \xi_2^2} +
{\Sigma^{(2)}(t_2,t_3) \over \bar \xi_2^2} +
{\Sigma^{(3)}(t_1,t_2,t_3) \over \bar \xi_2^3}= \exp \chi_3,
\end{equation}
where
\begin{equation}
\chi_3={ \Psi(t_1) \over \bar \xi_2} + { \Psi(t_2) \over \bar
\xi_2} + { \Psi(t_2)\over \bar \xi_2} + {\Psi^{(2)}(t_1, t_2)
\over \bar \xi_2^2} + {\Psi^{(2)}(t_2, t_3) \over \bar \xi_2^2} +
{\Psi^{(2)}(t_1, t_3) \over \bar \xi_2^2} + {\Psi^{(3)}(t_1, t_2,
t_3) \over \bar \xi_2^3}
\end{equation}
and
\begin{equation}
{\Psi^{(3)}(t_1, t_2, t_3) \over \bar \xi_2^3} =
\sum {t_1^p t_2^q t_3^r \over p! q! r!} {\langle \delta^p(x_1)\delta^q(x_2)\delta^r(x_3)
\rangle \over {\bar \xi_2}^{(p+q+r)}}
\end{equation}
The quantity $\Sigma^{(3)}$ is similarly the three-point
generalisation of the two-point quantity $\Sigma^{(2)}$. Expanding
this relation one can express three-point moments in terms of
factorial moments of the three-point CPDF.
\begin{eqnarray}
\Psi^{(3)}_{111} & = & ( 2- 3 \Sigma^{(2)}_{11} +
\Sigma^{(3)}_{111})
\\ \Psi^{(3)}_{112} & = & {1 \over {\bar \xi}_2}(-3 -
{\Sigma^{(2)}_{11}}^2 - \Sigma^{(3)}_{111} + {\bar \xi}_2
\Sigma^{(3)}_{112} - 2{\bar \xi}_2 \Sigma^{(2)}_{12} + 2{\bar
\xi}_2 \Sigma_2 + \Sigma^{(2)}_{11}(5 - \bar \xi_2 \Sigma_2)) \\
\nonumber \Psi^{(3)}_{113} & = & {1 \over {\bar \xi}_2^2}( 4 +
2\Sigma_{11}^2 + \Sigma^{(3)}_{111} + 4 \xi_2 \Sigma^{(2)}_{12} -
2\bar \xi_2^2 \Sigma^{(2)}_{13} - 6 \bar \xi_2 \Sigma_2 - \bar
\xi_2 \Sigma^{(3)}_{111}\Sigma_2 \\ \nonumber & &-\bar \xi_2
\Sigma^{(3)}_{112} + 2 \bar \xi_2^2 \Sigma_3 - \Sigma^{(2)}_{11}(
7 + 2 \xi_2( \Sigma^{(2)}_{12} - 3 \Sigma_2) + \bar \xi_2^2
\Sigma_{3}) + {\bar \xi_2}^2 \Sigma_{113})
\end{eqnarray}

It is possible to generalise these equations for an arbitrary
number of points:
\begin{equation}
1 +  \sum {\Sigma(t_i)\over \bar \xi_2^2} + \sum_{\rm pairs}
{\Sigma^{(2)}(t_i, t_j)\over \bar \xi_2^2} + \sum_{\rm triplets}
{\Sigma^{(3)}(t_i, t_j, t_k)\over \bar \xi_2^2} + \dots +
{\Sigma^{(p)}(t_i \dots t_n) \over \bar \xi_2^p} = \exp
\chi_p,\label{gen1}
\end{equation}
where
\begin{equation} \chi_p =  \sum {\Psi(t_i) \over \bar \xi_2}
+ \sum_{\rm pairs} {\Psi^{(2)}(t_i, t_j) \over \bar \xi_2^2} +
\sum_{\rm triplets} {\Psi^{(3)}(t_i, t_j, t_k) \over \bar \xi_2^3}
\dots + {\Psi^{(p)}(t_1, t_2, \dots t_k) \over \bar \xi_2^p}.
\end{equation}
It is important to note that two-point quantities depend only on
the separation of two cells, but the multiple-point  correlators
and their generating functions depend on the geometrical
configuration of different cells. We have used the following
definition for multiple-point cumulant correlators:
\begin{equation}
{\Psi^{(k)}(t_1, t_2, \dots t_k) \over \bar \xi_2^k} = \sum \left(
{t_1^p t_2^q t_3^r \dots t_k^s \over p! q! r! \dots s!}\right)
{\langle \delta^p(x_1)\delta^q(x_2)\delta^r(x_3)\dots
\delta^s(x_k) \rangle \over {\bar \xi_2^{(p+q+r+ \dots+s)} }}
\end{equation}
Inverting this relation, we can express  $\Psi^{(k)}(t_1, t_2,
\dots, t_k)$ in terms of the generating function for factorial
moments, which can be measured directly from simulations:
\begin{eqnarray}
\Psi^{(k)}(t_1, t_2, \dots, t_k) &= & \ln M_k(t_1, t_2, \dots,
t_k) - \sum_{(k-1) \rm tuplets}  \ln M_{k-1}(t_1, t_2, \dots,
t_{k-1}) +  \sum_{(k-2) \rm tuplets}  M_{k-2}(t_1, t_2, \dots,
t_{k-2}) + \\ \nonumber && \dots  + (-1)^{(k-2)} \sum_{\rm pairs}
\ln M_2(t_i, t_j) + (-1)^{(k-1)} \sum_{\rm singlets} \ln M(t_i)
\end{eqnarray}
where we have defined
\begin{eqnarray}
M(t_1) & = &  1 + {\Sigma(t_1) \over \xi_2} \\ M_2(t_1,t_2) & = &
1 + {\Sigma(t_1) \over \xi_2} + {\Sigma(t_2) \over \xi_2} +
{\Sigma^{(2)}(t_1,t_2) \over \xi_2^2}.
\end{eqnarray}
It is also possible to work with factorial correlators instead of
the factorial moments defined above by extending the two-point
factorial correlators introduced by Szapudi \& Szalay (1995) to
multiple points. If we denote the $k^{th}$ order factorial moments
in different cells by $(N_i)_k$ then we can write the generating
function of the two-point factorial correlators $W_{kl}$ by
$W(t_1, t_2)$, the generating function of the three-point
factorial correlator $W_{klm}$ by  $W(t_1, t_2, t_3)$, and so on.
In general $W(t_1, \dots, t_s)$ denotes the generating function
for s-point factorial correlators $W_{k \dots r}$:
\begin{eqnarray}
W^{(1)}( t_1 ) & = & 1 + \sum_{k=1}^{\infty} {1 \over k!} \Big (
{t_1 \over {\bar N}} \Big )^k \langle (N_1)_k \rangle\\
W^{(2)}(t_1, t_2) & = & \sum_{k=1, l=1}^{\infty} {1 \over k! l!}
\Big ( {t_1 \over {\bar N}} \Big )^{k} \Big ( {t_2 \over {\bar N}}
\Big )^{l}\langle ((N_1)_k - \langle (N_1)_k \rangle) ((N_2)_l -
\langle (N_2)_l \rangle)  \rangle  \\ W^{(3)}(t_1, t_2,t_3) & = &
\sum_{k=1, l=1, m=1}^{\infty}  {1 \over k! l! m!}  \Big ( {t_1
\over {\bar N}} \Big )^{k} \Big ( {t_2 \over {\bar N}} \Big )^{l}
\Big ( {t_3 \over {\bar N}} \Big )^{m} \langle ((N_1)_k - \langle
(N_1)_k \rangle) ((N_2)_l - \langle (N_2)_l \rangle)((N_2)_m -
\langle (N_2)_m \rangle)  \rangle \\ W^{(s)}(t_1, \dots ,t_s) & =
& \sum_{k=1, \dots , r=1}^{\infty}  {1 \over k! \dots r!}  \Big (
{t_1 \over {\bar N}} \Big )^{k} \dots \Big ( {t_s \over {\bar N}}
\Big )^{r} \langle ((N_1)_k - \langle (N_1)_k \rangle) \dots
((N_s)_r - \langle (N_s)_r \rangle)  \rangle
\end{eqnarray}
 Now it is possible to link the basis function  $W^{(n)}$ for computing multi-point cumulant correlators
 with the other set of basis functions $\Sigma^{(n)}$, which carries equivalent information.
 \begin{eqnarray}
W \Big ( { t_1 \over \bar \xi_2} \Big )  & = & 1 + { \Sigma(t)
\over \bar \xi_2} \\ W^{(2)} \Big ( { t_1 \over \bar \xi_2}, { t_2
\over \bar \xi_2} \Big ) & = & { 1 \over \xi_2^2 }\Sigma^{(2)}
(t_1, t_2) - { 1 \over \xi_2} \Sigma(t_1) { 1 \over \xi_2}
\Sigma(t_2)
\\  \nonumber
W^{(3)} \Big ( { t_1 \over \bar \xi_2}, { t_2 \over \bar \xi_2}, {
t_3 \over \bar \xi_2} \Big ) & = & { 1 \over \xi_2^3 }\Sigma^{(3)}
(t_1, t_2, t_3) + 2 { 1 \over \xi_2} \Sigma(t_1) { 1 \over \xi_2}
\Sigma(t_2) { 1 \over \xi_2} \Sigma(t_3)
\\
& & - { 1 \over \xi_2} \Sigma(t_1) { 1 \over \xi_2^2}
\Sigma^{(2)}(t_2, t_3)  - { 1 \over \xi_2} \Sigma(t_2) { 1 \over
\xi_2^2} \Sigma^{(2)}(t_1, t_3) - { 1 \over \xi_2} \Sigma(t_3) { 1
\over \xi_2^2} \Sigma^{(2)}(t_1, t_2)
\end{eqnarray}
Relating multi-point factorial moments with generating functions
of multiple-point cumulant correlators (defined earlier) we can
write:
\begin{eqnarray}
\exp (\chi_1) &=& \exp \Big ( { \Psi(t_1) \over \xi_2} \Big ) = W
\Big ( { t \over \bar \xi_2} \Big ); \\ \exp (\chi_2) & = &\exp
\left( {\Psi(t_1) \over \xi_2} +  { \Psi(t_2) \over \xi_2} +
{\Psi^{(2)}(t_1, t_2) \over \xi_2^2} \right)  =  W^{(2)} \Big
({t_1 \over \bar \xi_2}, {t_2 \over \bar \xi_2} \Big ) +  W \Big
({t_1 \over \bar \xi_2} \Big ) W \Big ({t_2 \over \bar \xi_2} \Big
)  \\ \exp (\chi_3) & = & W^{(3)} \Big ({t_1 \over \bar \xi_2}, {t_2
\over \bar \xi_2},{t_3 \over \bar \xi_2} \Big ) \\ && +  W^{(2)}
\Big ({t_1 \over \bar \xi_2}, {t_2 \over \bar \xi_2} \Big ) W \Big
({t_3 \over \bar \xi_2} \Big ) +  W^{(2)} \Big ({t_1 \over \bar
\xi_2},{t_3 \over \bar \xi_2} \Big ) W \Big ({t_2 \over \bar
\xi_2}\Big ) +  W^{(2)} \Big ({t_2 \over \bar \xi_2},{t_3 \over
\bar \xi_2} \Big ) W \Big ({t_1 \over \bar \xi_2} \Big ) + W \Big
({t_1 \over \bar \xi_2} \Big ) W \Big ({t_2 \over \bar \xi_2} \Big
) W \Big ({t_3 \over \bar \xi_2} \Big )
\end{eqnarray}
Generalizing these results, and connecting multi-point factorial
correlators with multi-point void probability functions we obtain
the result equivalent to eq(\ref{gen1}).
\begin{eqnarray}
&&W^{(s)} \Big ({t_1 \over \bar \xi_2}, \dots ,{t_{s} \over \bar
\xi_2} \Big ) + \sum_{\rm singlets}  W^{(1)} \Big ({t_i \over \bar
\xi_2} \Big )  W^{(s-1)} \Big ({t_1 \over \bar \xi_2},
\dots,{t_{s-1} \over \bar \xi_2} \Big ) + \sum_{\rm pairs}  W \Big
({t_i \over \bar \xi_2} \Big ) W \Big ({t_j \over \bar \xi_2} \Big
) W^{(s-2)} \Big ({t_1 \over \bar \xi_2}, \dots ,{t_{s-2} \over
\bar \xi_2} \Big )
\\ \nonumber
&&+ \dots + \sum_{\rm (s-2)tuplets}  W \Big ({t_1
 \over \bar \xi_2} \Big ) \dots W \Big ({t_{s-2} \over \bar \xi_2} \Big ) W^{(2)}
 \Big ({t_i \over \bar \xi_2},{t_j \over \bar \xi_2} \Big ) +
 W \Big ({t_i \over \bar \xi_2} \Big ) \dots W \Big ({t_s \over \bar \xi_2}
 \Big )\\ \nonumber
&&= \exp \Big ( \sum {\Psi(t_i) \over \bar \xi_2} + \sum_{\rm pairs} {\Psi^{(2)}(t_i, t_j)
\over \bar \xi_2^2} + \sum_{\rm triplets} {\Psi^{(3)}(t_i, t_j, t_k) \over \bar \xi_2^3}
\dots + {\Psi^{(p)}(t_1, t_2, \dots t_k) \over \bar \xi_2^p}\Big ).
\end{eqnarray}
It is to be noted that multi-point cumulant correlators can also
be defined in such a way that they incorporate all lower order
correlations. For example, if we allow
 one or more of the powers $p,q,r$ or $s$ to vanish the correlator becomes independent
of one or more spatial co-ordinates. This will allows a more
compact way to express factorial moments in terms of multi-point
cumulant correlators (Szapudi \& Szalay 1997) which is equivalent
to the result obtained using factorial moments. However, it is useful to
decompose multi-point factorial moments and multi-point factorial
correlators in terms of irreducible one-point and two-point
quantities from a computational point of view. As we shall see,
this also allows us to relate multi-point cumulant correlators
directly with the tree amplitudes involved in the hierarchical
ansatz.

The discussion so far is completely general. While Szapudi \&
Szalay (1993) used the hierarchical approximation of $Q_{N+M} =
Q_NQ_M$ and analogous approximations for higher orders, we will be
concentrating on the formalism developed by Bernardeau \&
Schaeffer (1992) in which amplitudes associated with hierarchical
trees are assumed to be the product of tree vertices. Following
paper I, we can now relate $\Psi^{(n)}(t_1, \dots, t_n)$ with
generating functions $\mu_n(t)$ of vertices appearing in
tree-representation of multi-point cumulant correlators which
depends on the hierarchical ansatz. These vertices depend on the
tree-level amplitudes $\nu_n$ in representation of correlation
hierarchy of matter correlation function (see paper I for complete
discussion):
\begin{eqnarray}
\mu_1(t) & = & \sum_{n=1} {C_{n1} t^n \over n!} \\ \mu_2(t) & = &
\sum_{n=1} {C_{n11} t^n \over n!} \\ \mu_3(t) & = & \sum_{n=1}
{C_{n111} t^n \over n!};
\end{eqnarray}
etc. The quantities $\mu_n$ determine the amplitude associated
with a vertex that has $n$ external legs. Using the vertex
generating function we can express $\Psi^{(n)}(t_1, \dots, t_n)$
as:
\begin{eqnarray} \Psi^{(2)}(t_1,t_2) & = & \mu_1(-t_1) \xi_{ab}
\mu_1(-t_1) \\ &&\Psi^{(3)}(t_1,t_2,t_3) = \mu_1(-t_1) \xi_{ab}
\mu_2(-t_1)\xi_{ac} \mu_1(-t_1) + \dots {\rm
(cyclic~permutations)} \\ \nonumber \Psi^{(3)}(t_1,t_2,t_3, t_4) &
= & \mu_1(-t_1) \xi_{ab} \mu_3(-t_1)\xi_{ac} \mu_1(-t_1)\xi_{ad}
\mu_1(-t_1) + \dots {\rm (cyclic~permutations)} \\ && +
\mu_1(-t_1) \xi_{ab} \mu_2(-t_1)\xi_{ac} \mu_2(-t_1)\xi_{ad}
\mu_1(-t_1) + \dots {\rm (cyclic~permutations)}
\end{eqnarray}
In deriving above relations we have considered only the dominant
contribution and this  approximation is valid only when the
density variance in each cell is much higher than the correlation
between cells. Szapudi \& Szalay (1997) proposed the use of the
two-point cumulant correlator to separate amplitudes associated
with different tree vertices, but since the number of tree
topologies increases exponentially with order of the diagram (Fry
1984), the two-point cumulant correlator can only be useful in up
to $4^{th}$ order diagrams. This deficiency can only be cured by
moving to multi-point cumulants and our formalism developed here
will be able to determine $\nu_n$ parameters for arbitrary $n$.
However, it is obvious that multi-point cumulants are in general
dependent on the geometric configuration of the points, and their
determination becomes more complicated as the order increases. We
also  stress the point that our formalism can in general be used
in principle to determine all the tree amplitudes without making
any assumptions at all. This is very useful as this can actually
test the validity of all hierarchical models of gravitational
clustering in the highly nonlinear regime.
\begin{figure*}
\protect\centerline{
\epsfysize = 4.truein
\epsfbox[0 0 473 584]
{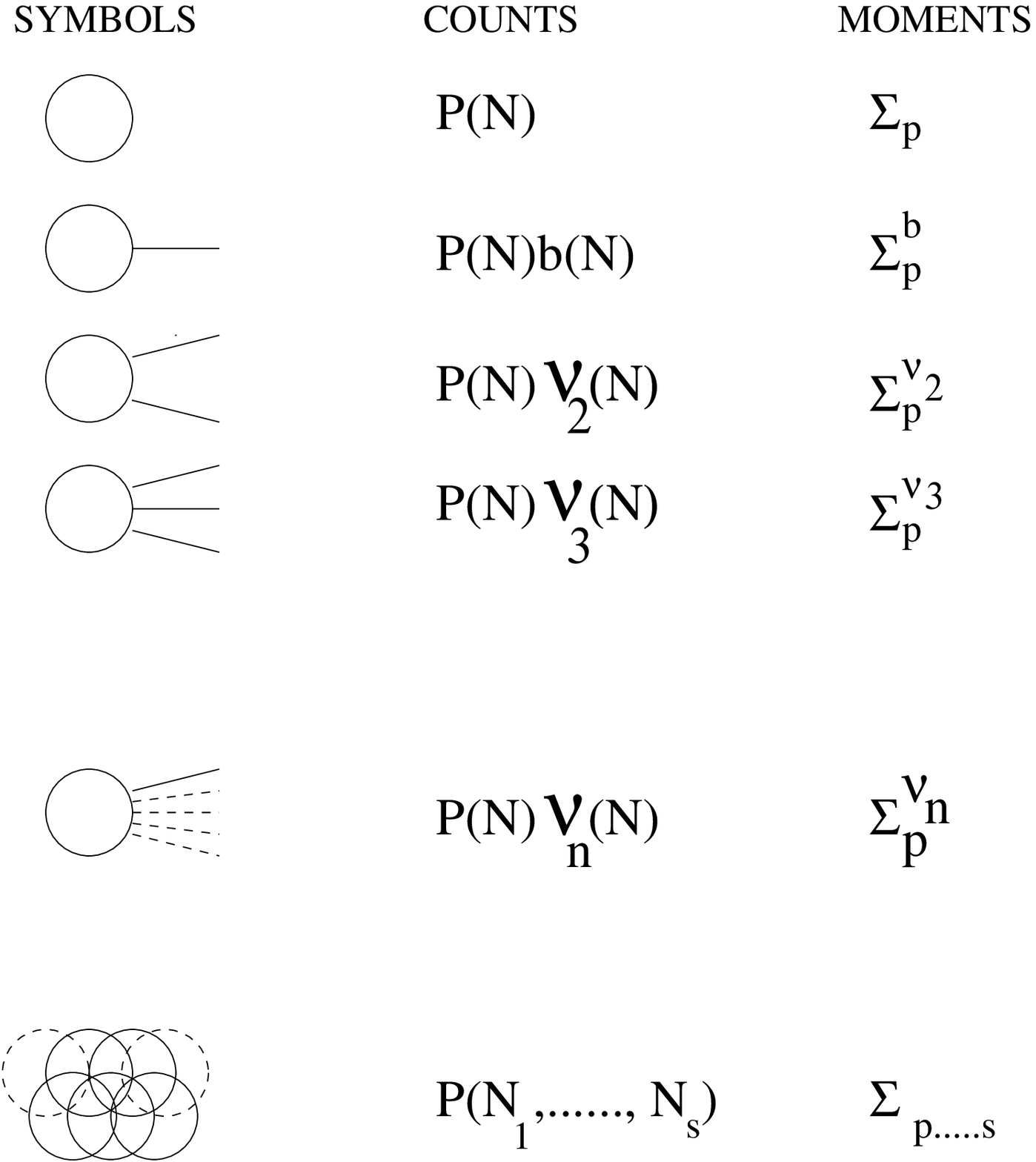} } \caption{Basic elements in the representation of
multi-point joint CPDF and related factorial moments. We represent
$P(N)$ and one associated factorial moment $\Sigma_p$ by a circle
with no legs; $P(N)b(N)$ and associated moments $\Sigma_p^b$ are
represented by circle with one leg and, similarly, a circle with
$n$ legs represents $P(N)\nu_n(N)$ and its associated factorial
moments $\Sigma_p^{\nu_n}$, where $p$ represents the order of the
relevant factorial moments, and $n$ the order of the relevant
vertex. }
\end{figure*}

\begin{figure*}
\protect\centerline{
\epsfysize = 4.truein
\epsfbox[0 0 745 560]
{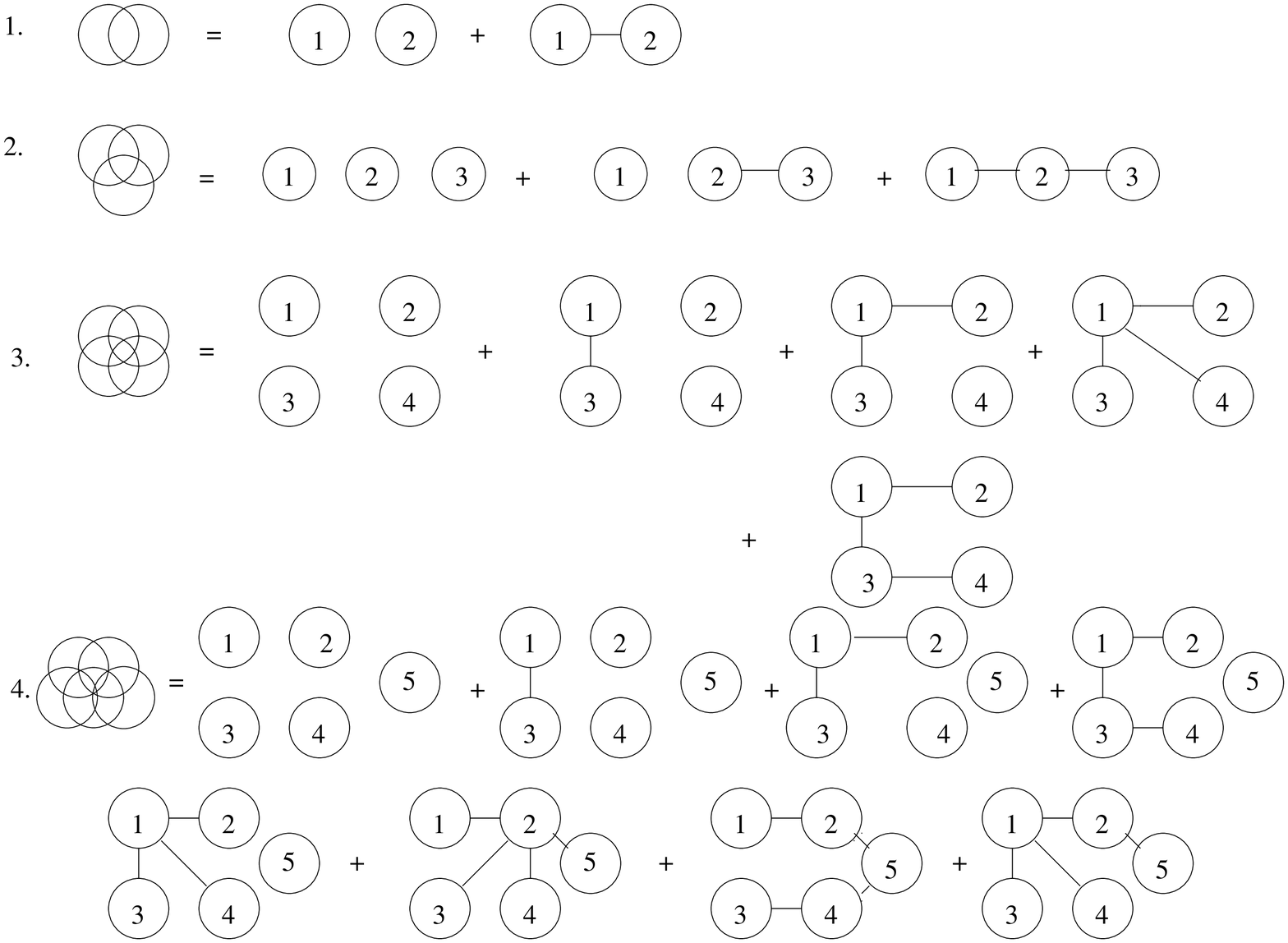} } \caption{The decomposition of the MCPDF is
represented diagrammatically. Every straight line joining two
spheres represents the correlation between two cells. Higher-order
terms contain more straight lines joining a given set of cells and
hence can be ignored in a leading-order approximation. Only tree
terms are shown. All possible permutations of labels are to be
considered, but in this figure we show only one diagram for each
topology. In second order in $\xi_{ij}/ \bar \xi_i$, it is
possible only to have one type of topology, in third order we can
have a two different topologies represented by  ``snake'' and
``star'' diagrams. In even higher order it is possible to have a
hybrid of these two different types. It is interesting to note
that the correlation hierarchy for matter induces exactly similar
hierarchy for MCPDF.}
\end{figure*}

The multi-point CPDF (MCPDF) can be decomposed in terms of several
one-point quantities. Expanding the multi-point void probability
function, which acts as a generating function for MCPDF,
Bernardeau \& Schaeffer(1992) showed that it is possible to
express the MCPDF in terms of a series expansion in
$\xi_{ij}/\bar{\xi}_2 << 1$, where $\xi_{ij}$ is the two-point
correlation function between different overdense cells and $\bar
\xi_2$ is the variance within one cell as defined above. Such a
decomposition is expected to work for overdense cells when they
are separated by moderately large distance. To first order in such
a decomposition it was found that
\begin{equation}
P(N_1, N_2) = P(N_1)P(N_2) +
P(N_1)b(N_1)\xi_{12}(r_{12})P(N_2)b(N_2),
\end{equation}
where $b(N_1)$ is the bias function for overdense cells which
describes how over-dense cells are correlated compared with
background mass. Normalisation constrains on $P(N_1, N_2)$ can be
translated in to constrains for $b(N)$.

\begin{eqnarray}
\sum b(N)P(N) & = & 0; \\ \nonumber \sum N b(N)P(N) &  =  &\bar N.
\end{eqnarray}
With such a decomposition we can express two-point quantities,
such as $\Sigma_{pq}$, in terms of one-point factorial moments but
with the effect of bias $b(N)$ taken into account:
\begin{equation}
\Sigma_{pq} = \Sigma_p\Sigma_q + \Sigma_p^b \xi_{12} \Sigma_q^b,
\end{equation}
where we have defined a new biased one point factorial moment $\Sigma_p^b$ by,
\begin{equation}
\Sigma_p^b = \sum {\xi_2 \over p! N_c^p}   N(N -1)\dots(N -p +1) b(N) P(N)
\end{equation}
A similar decomposition can be performed for the 3-point CPDF and,
if we write the expression up to second order in $\xi_{ij} \bar
\xi_2$, we get
\begin{eqnarray}
P(N_1, N_2, N_3)  & =  & P(N_1)P(N_2)P(N_3) \\ \nonumber & & +
P(N_1)P(N_2)b(N_2)\xi_{23}P(N_3)b(N_3) +
P(N_2)P(N_3)b(N_3)\xi_{31}P(N_1)b(N_1) + P(N_3)P(N_1)b(N_1)
\xi_{12}P(N_2)b(N_2) \\ \nonumber & &+\nu_2(N_1)P(N_1)\xi_{12}
b(N_2)P(N_2)\xi_{13}b(N_3)P(N_3) +\nu_2(N_2)P(N_2)\xi_{23}
b(N_3)P(N_3)\xi_{21}b(N_1)P(N_1) \\ \nonumber
&&+\nu_2(N_3)P(N_3)\xi_{13} b(N_1)P(N_1)\xi_{32}b(N_2)P(N_2).
\end{eqnarray}
A new function appears, together with its associated constraints:
\begin{eqnarray}
\sum \nu_2(N)P(N) & = & 0 \\ \nonumber \sum N \nu_2(N)P(N) & = &
\nu_2 {\bar N};
\end{eqnarray}
$\nu_2$ is the second-order vertex associated with a
tree-development of the correlation hierarchy in the highly
non-linear regime. We define the associated one-point factorial
moment $\Sigma_p^{\nu}$, which takes into account of corrections
second order in $\xi_{ij}\bar \xi_2$:
\begin{equation}
\Sigma_p^{\nu_2} = \sum {\xi_2 \over p! N_c^p} N(N -1)\dots(N -p
+1) \nu_2(N_1) P(N_1).
\end{equation}
With this higher order contributions are taken into account
$\Sigma_{pqr}$ can be expressed in terms of one point quantities
such as $\Sigma_p, \Sigma_q^b~\rm~ and ~\Sigma_q^{\nu_2}$,
\begin{equation}
\Sigma_{pqr} = \Sigma_p \Sigma_q \Sigma_r + \Sigma_p\Sigma_q^b \xi_{23}\Sigma_r^b
+ \Sigma_q \Sigma_p^b\xi_{13}\Sigma_r^b + \Sigma_r \Sigma_p^b \xi_{12} \Sigma_q^b
+ \Sigma^{\nu_2}_q\xi_{23}\Sigma_p^b\xi_{13}\Sigma_r^b + \Sigma^{\nu_2}_r \xi_{31}\Sigma_p^b \xi_{12}\Sigma_q^b
 + \Sigma^{\nu_2}_p \xi_{31}\Sigma_p^b \xi_{32}\Sigma_q^b
\end{equation}
It was shown in paper I that such a decomposition is possible to
arbitrary order, as long as we neglect the contributions from loop
terms. The tree structure of MPCDF and associated multi-point
factorial moments  is induced by the inherent tree structure of
correlation functions and is very similar in nature. It is
interesting to note that, in this particular case, contributions
from different orders are completely separated and also the
contribution to joint moments from different spatial locations can
be expressed as different independent local terms depending only
on one spatial co-ordinate. This also makes volume corrections
easier to perform, as we shall see in \S 4.

\section{Scaling in One-Point and Two-Point Quantities}
Scaling properties that arise because of the inherent tree
structure of correlation functions in highly non-linear regime
were studied by Balian \& Schaeffer (1989). The principal
assumption in this and related studies is that the  vertex
amplitudes which appear in a tree development of higher-order
correlation functions should be constants. Every higher-order
correlation function is build by taking suitable products of
two-point correlation functions and all such $N$ different tree's
carry different amplitudes:
\begin{equation}
\xi_N( {\bf r}_1, \dots {\bf r}_N ) = \sum_{\alpha, \rm N-trees} T_{N,\alpha}
\sum_{\rm labelings}
\prod_{\rm edges}^{(N-1)} \xi_{(r_i, r_j)}.
\end{equation}
which in most general case can be expressed as
\begin{equation}
\xi_N( \lambda {\bf r}_1, \dots  \lambda {\bf r}_N ) = \lambda^{-\gamma(N-1)}
\xi_N( {\bf r}_1, \dots {\bf r}_N ).
\end{equation}
Clearly no loop terms (such as those that appear in the Kirkwood
scaling relation) are considered in such an analysis. There is
some support from observations for such an assumption in highly
nonlinear regime. On the other hand, in the quasi-linear regime
and in the limit of vanishing variance such a hierarchy occurs
naturally but the hierarchical coefficients in general depend on
different shape parameters appearing in higher-order correlations.
Recent studies of the bispectrum, however, show that
nonlinearity tends to smooth out such shape dependence and in the
highly non-linear regime the bispectrum becomes completely
independent of shape. Such a result of course is in agreement with
the hierarchical ansatz where all higher-order parameters are
assumed to become shape independent (Scoccimarro et al. 1998).

Any assumption regarding the entire hierarchy of many-body
correlation functions has very strong consequences and can be used
to derive properties of one-point and multi-point statistics.  It is
important to note that since the exact values of hierarchical
coefficients are left unspecified in such an ansatz, it is only
possible to talk about generic scaling properties induced by them.

The void probability function (VPF), which is very simple to measure
from N-body data or galaxy catalogs carries information on volume
averages of all orders of correlation functions. It has been shown
that VPF $P_V(0)$ which denotes the probability of finding a cell
of volume $V$ empty of any particle  can be expressed as an unique
function of scaling variable $N_c$, defined above in \S 2(White
1979; Balian \& Schaeffer 1989 ). In particular we have (Balian \&
Schaeffer 1989)
\begin{equation}
P_V(0) = \exp \left( -{\bar N} \sigma(N_c)\right) = \exp \left( -
{\phi(N_c) \over \bar \xi_2}\right) = \exp\left( {\Psi(-N_c) \over
\bar \xi_2}\right)
\end{equation}
For large values of $N_c$, $\sigma(N_c)$ becomes a power-law with
power law index $-\omega$ that depends on initial conditions
\begin{equation}
\sigma(N_c) = a N_c^{-\omega}
\end{equation}
(Balian Schaeffer (1989). Note that the VPF depends by
construction on the discrete nature of sampling and it is not
possible to define a VPF for a continuous matter distribution. In
effect, however, the VPF can be used to learn about the $S_N$
parameters of the underlying field.

Scaling in VPF induces a similar scaling in the CPDF, and the
evolution of the CPDF through different length-scales and
different epochs can be described using a unique scaling function
$h(x)$ (with $x = N/N_c$). This function is related to the VPF
through
\begin{equation}
h(x) = - \int_{-i\infty}^{i\infty} { dy \over 2 \pi i} \phi(y) \exp(yx)
\end{equation}
(Balian \& Schaeffer 1989) and $P_V(N)$ is related to $h(x)$ by
the following expression (Balian \& Schaeffer 1989),
\begin{equation}
P_V(N) = { 1 \over N_c {\bar \xi_2}} h \left( { N \over N_c}
\right)
\end{equation}
Using these relations it is now possible to express $P_V(N)$ as
\begin{equation} P_V(N) = { a \over {\bar \xi_2} N_c} { 1 -
\omega \over \Gamma(\omega)} \left( {N \over N_c} \right)^{\omega
- 2}
\end{equation}
 (Balian \& Schaeffer 1989). This expression is valid in the range $N_v<N<N_c$ where, physically, $N_v$ is
the typical cell occupancy in underdense regions and $N_c$ in
overdense regions; these are related to each other by $N_v = N_c
(a/\xi_2)^{1/(1-\omega)}$. It is interesting to note that the
power-law index $-\omega$,  which appears in the expression of the
VPF, also appears in case of CPDF.

For highly overdense cells a different asymptote can be computed
by noting the that $\phi(y)$ exhibits a singularity for small
negative values of $y=y_s$
\begin{equation}
\phi(y) = \phi_s-a_s\,\Gamma(\omega_s)\,(y-y_s)^{-\omega_s}
\end{equation}
It introduces an exponential cut-off for large values of $N$ in $P(N)$ (Balian \& Schaeffer, 1989),
\begin{equation}
h(x) = a_s\,x^{\omega_s-2}\exp(-\vert y_s\vert\,x),
\end{equation}

Previous studies have shown that an analytical fitting function
for $h(x)$ works well  in both these regimes (Balian \& Schaeffer
1989):
\begin{equation} h(x) = {a ( 1 - \omega) \over
\Gamma(\omega)} { x ^{\omega - 2} \exp(-x|y_s|) \over (1 + bx)^c}
\end{equation}
The constraints satisfied by $h(x)$ are
\begin{equation}
\int_0^{\infty} x h(x) dx = 1 \end{equation} and and
\begin{equation}
\int_0^{\infty} x^2 h(x) dx = 1.
\end{equation}

The hierarchical ansatz has also been used to predict a scaling
behaviour of the  multi-point CPDF and related statistics. The
formalism developed by Bernardeau \& Schaeffer (1992) shows that
the two-point CPDF can be written as
\begin{equation}
P_V(N_1, N_2) = P(N_1)P(N_2)\left[ 1 + \xi_2(r_1,
r_2)b(N_1)b(N_2)\right],
\end{equation}
where we have denoted joint occupation probability of two
different cells (we take sizes of these two cells to be the same)
by $P_V(N_1, N_2)$. Such a decomposition means that the
correlation function of two cells with occupancy $N_1$ and $N_2$
(Bernardeau \& Schaeffer 1992) is
\begin{equation}
\xi_2(N_1,N_2) = b(N_1)b(N_2) \xi_2(r_1, r_2).
\end{equation}

Several characteristic features of the bias induced by these
scaling properties can be noted. For one thing, it is clear from
this derivation that the bias  depends only on $x$ and hence only
on intrinsic properties of the collapsed object, at least for
large separations. Since the bias $b(N)$ satisfies the property
$\sum_{N}P(N)b(N) = 0$ it is clear that if the bias for some
occupation numbers is positive, it must be negative for others.
Detailed calculations show that cells with occupation number close
to $N_v$ (the average occupation of cells in underdense regions)
have negative bias. On the other hand, the correlation of two
identical objects (i.e. cells with same occupation number) is
always positive:
\begin{equation}
b(x) = \Big ( { \omega \over 2 a} \Big )^{1/2} { \Gamma[\omega] \over \Gamma[{1 \over 2}
 ( 1 + \omega)]} {x ^{(1 - \omega) / 2}}; ~~~~~~({\rm for}~~~~~ N_v < N < N_c)
\end{equation}
\begin{equation}
b(x) = { x \over x_{bs}}; ~~~~~~({\rm for}~~~~~ N > N_c)
\end{equation}
Interesting though $b(x)$ is in itself, from a computational point
of view it is  much easier to compute $b(>x)$, the bias of
overdense cells above certain threshold, defined as
\begin{equation}
b(>x) = {\int_x^{\infty} b(x')h(x')dx' \over  \int_x^{\infty} h(x')dx'}.
\end{equation}
It can be shown in the two regimes we have discussed that $b(>x)$
can be related directly with $b(x)$, i.e. $b(>x) = 2b(x)$ for
$N<N_c$ and $b(>x) = b(x)$ for $N>N_c$. It is interesting to point
out at this point that for a Gaussian density field $b(>x) = b(x)
= 1$ for rare events. In the quasilinear regime, the value of $b(
>x)$ computed numerically has been shown to match theoretical predictions
(Bernardeau 1996). However no such studies have been made in
the highly nonlinear regime where we expect the hierarchical
ansatz to be valid.

In paper I we derived higher-order predictions of the hierarchical
ansatz, but we restrict ourselves here to two-point quantities
associated with overdense cells. The  bias in this case acts as a
generating function for the cumulant correlators; this was studied
in Munshi \& Melott (1998). A complete test of higher-order
predictions covering the  CPDF and VPF of overdense cells will be
presented elsewhere.

\section{Simulations and Data Analysis }

\begin{figure*}
\protect\centerline{
\epsfysize = 5.truein \epsfbox[20 146 587 714]{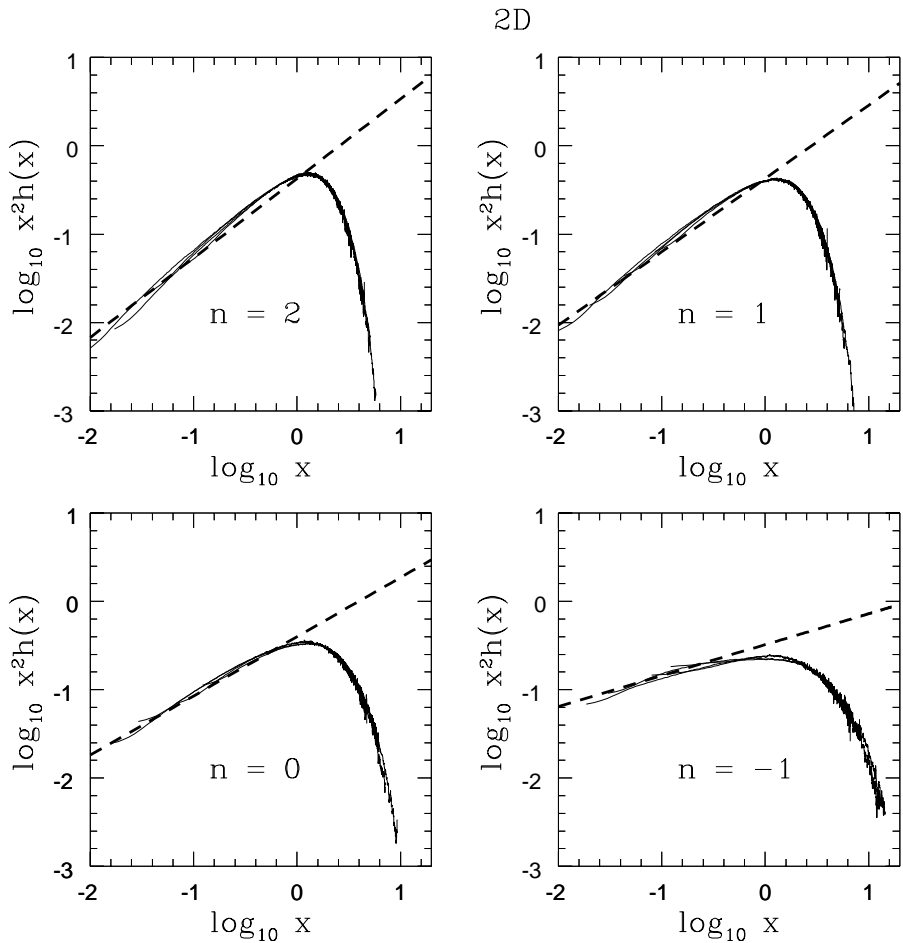} }
\caption{The measured $h(x)$ in 2D numerical simulations, as
described in the text. The large dynamic range allows us to
determine the function $h(x)$ to an accuracy at the level of
$10^{-3}$, and order-of-magnitude increase compared to all
previous studies in 2D or 3D. The function $h(x)$ in general shows
a power law profile for $x<1$ which is followed by an exponential
cutoff for $x>1$. This exponential cutoff is more severe when
there are more power at smaller scales. Since $x = N/N_c$ and
$N_c$ is the typical occupation of cells inside clusters, this
means that the CPDF $P_V(N)$ at any scale will have a power-law
profile for $N<N_c$ and will have an exponential tail for $N>N_c$.
In an infinite catalogue this exponential tail will extend to
infinitely large values of $N$. However, in a finite catalogue,
$P_l(N)$ shows an abrupt cutoff at $N_{\rm max}$, due to the
absence of any denser cells beyond that limit. Different length
scales map different regions of $h(x)$ curve. Previous studies
have shown that effect of finite volume corrections increases with
power on larger scales which reduces the available range of $x$
for which $h(x)$ can be measured accurately. The measured $h(x)$
satisfies two integral constraints ($S_1 = S_2 = 1$) and the power
law index of $x^2h(x)$ should be exactly the same as the negative
power-law index appearing in $\sigma(N_c)$.}
\end{figure*}

\begin{figure*}
\protect\centerline{
\epsfysize = 5.truein
\epsfbox[20 146 587 714]
{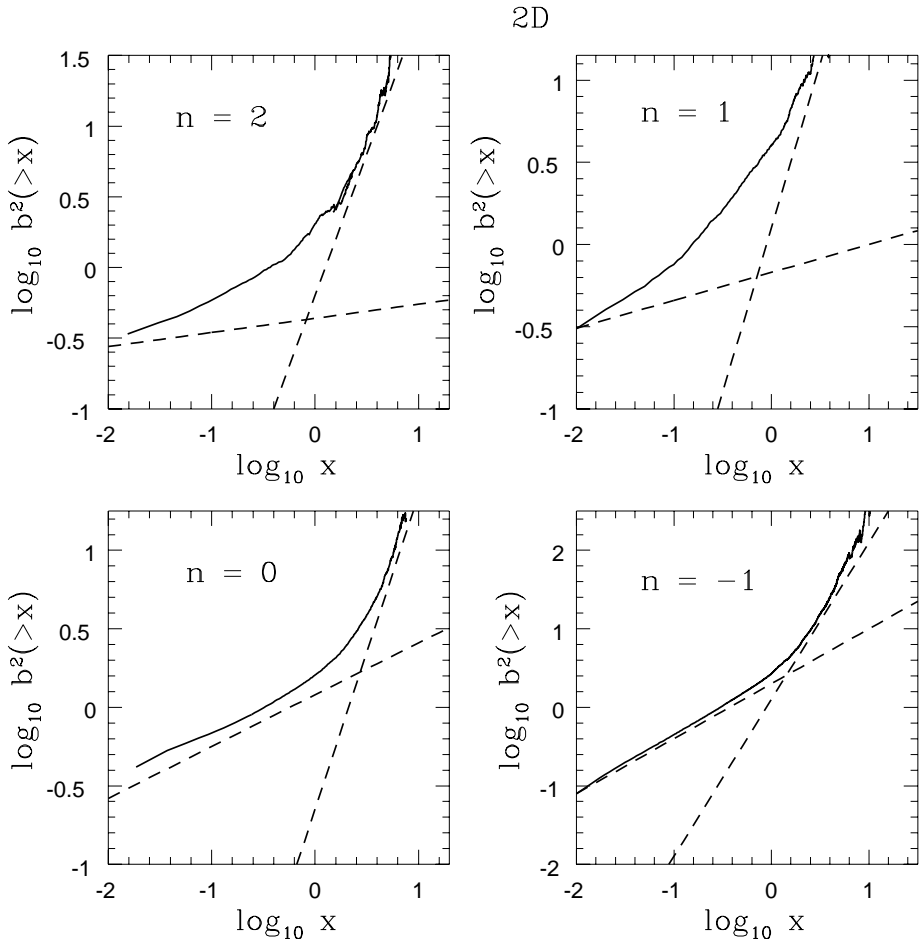} } \caption{The measured values of $b(>x)$ from the 2D
simulations described in the text. The bias $b(>x)$ shows two
different regimes similar to $h(x)$: for small values of $x$, it
increases slowly with $x$ according to a power-law with index less
than unity. This index is also related to the negative slope of
$\sigma(N_c)$ and slope of $x^2h(x)$ for $x<1$. For larger values
of $x$, $b(>x)$ shows a steeper increase, proportional to $x$. For
$b(N)$ this  means that moderately overdense cells show a slow
increase of bias with cell occupancy $N$ but for highly overdense
cells the bias increases linearly. The transition occurs for $N =
N_c$ which is the typical occupancy of cells in over-dense
regions. We have used two related but different methods to
estimate bias. The first method is based on measuring
cross-correlation of different classes of objects with respect to
background mass distribution and the second method is based on
computation of direct measurement of correlation function against
the background mass distribution for different classes of objects.
Good agreement between these two methods proves that the
factorization property of bias and also shows that the bias for an
overdense cell depends only on the intrinsic property of the
object concerned and can be expressed as a unique function of $x$.
}
\end{figure*}

\begin{figure*}
\protect\centerline{
\epsfysize = 5.truein
\epsfbox[20 146 587 714]
{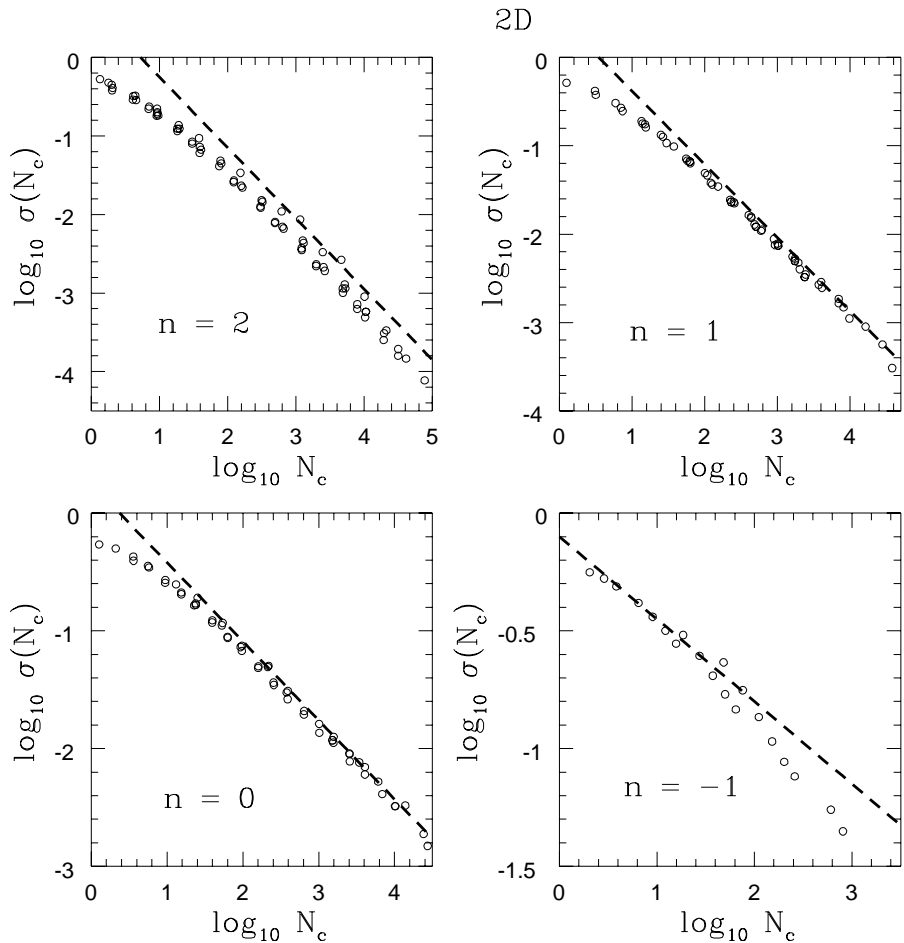} } \caption{The measured values of $\sigma(N_c)$ from
2D simulations described in the text. For a given cell size, the
parameter $N_c$ increases as the system evolves and for a given
epoch, the larger cells have higher values of $N_c$. Cell size we
analyze ranges from $(4l_{\rm grid})^2$ to $(64l_{\rm grid})^2$.
We have used different epochs of nonlinearity to construct the
function $\sigma(N_c)$. We also use different levels of
nonlinearity to increase available range of $N_c$. Different
dilution levels of the $4096^2$ data, i.e. $2048^2$, $1024^2$,
$512^2$ and $256^2$ were used to test the validity of the scaling
ansatz. For power spectra $n=2,1$ and $0$ we do not detect any
deviation from scaling models, but for $n=-2$ there are
discernible departures. This could be due to various spurious
effects which could influence the determination of $\sigma(N_c)$,
especially the initial grid which survives undistorted in
underdense regions if the initial spectrum is $n=-2$. Error
estimation in determination of $\sigma(N_c)$ was done by Colombi
et al. (1995) who found that the error is proportional to
$S_4$ which increases with more power on larger scales. However
all our data points satisfy the criteria $P_0 > 1/e$ for avoiding
errors introduced by the computational grid. This restricts the
maximum cell size for a given epoch for which $\sigma(N_c)$ can be
computed. The other possibility for violation of particular
scaling model which we have considered here arises from the fact
that instead of $S_N$ parameters being constant at highly
nonlinear regime they might show a very slow but steady increase
with level of nonlinearity as reported by Lucchin et. al (1994) and
Colombi et. al (1996).
However note that such an argument will also mean a clear
departure of CPDF from scaling which we have not detected in our
studies.}
\end{figure*}

\begin{figure*}
\protect\centerline{
\epsfysize = 2.75truein
\epsfbox[20 407 588 715]
{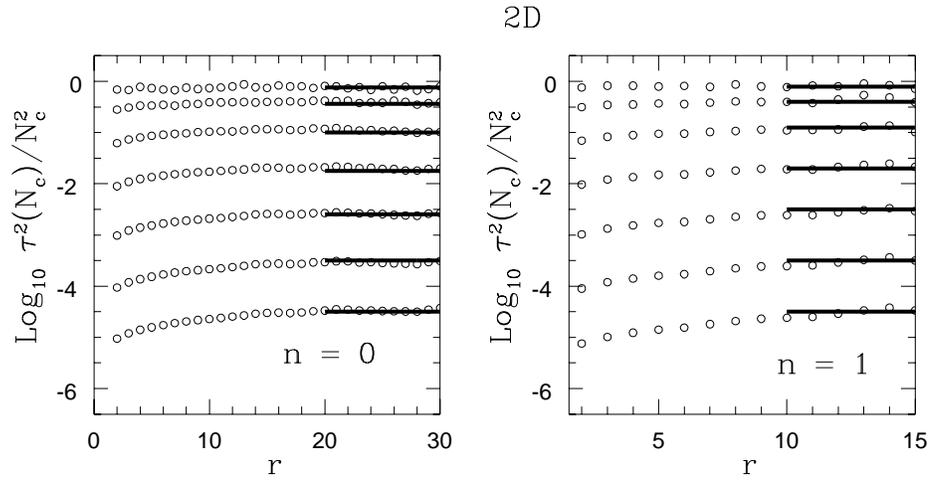} } \caption{ We plot the quantity $\tau^2(N_c)/N_c^2$
described in the text for different cell separations $r$. We have
shown results for only two different initial conditions for one
particular cell size of $(4l_{\rm grid})^2$. For each cell size we
consider different levels of dilutions. The lowest curve in each
panel correspond to $4096^2$ data. Each subsequent curve from
bottom to top correspond to a dilution by a factor of $1/4$ in
$\bar N$,  while the top-most curve correspond to $64^2$ data. The
solid lines at larger values of $r$ correspond to values of
$\tau^2(N_c)/N_c^2$ taken for scaling studies; this quantity
exhibits scaling properties when plotted as a function of $N_c$ as
shown in Figure 7. Departure from linear behaviour for small
separation is due to terms higher order in $\xi_{ij}/ \bar \xi_i$
which are neglected in  our study of the 2VPF. }
\end{figure*}

\begin{figure*}
\protect\centerline{
\epsfysize = 5.truein
\epsfbox[20 146 587 714]
{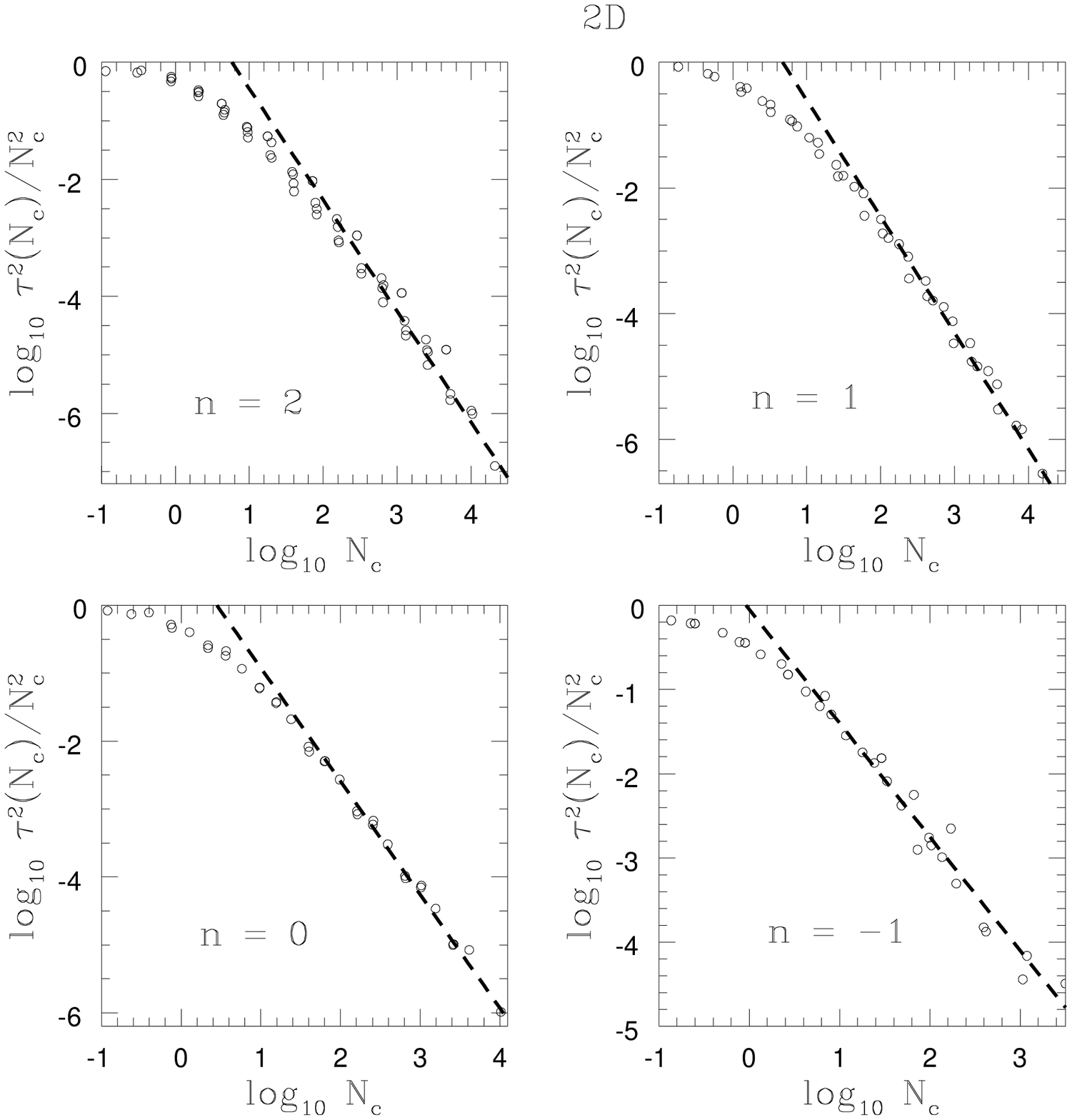} } \caption{The measured values of $\tau^2(N_c)/
N_c^2$ from 2D simulations described in the text. We compute the
2VPF $P(0,0)$ and VPF $P(0)$ using a regular grid for different
separation as explained in previous figures and use this to
estimate $\tau^2(N_c)/N_c^2$ for different values of $N_c$ in the
limit of large separations as described in the text. As in the
case of VPF we repeat the process for different values of cell
size from $(4l_{\rm grid})^2$ to $(64l_{\rm grid})^2$. We also use
different levels of dilution to increase the range of $N_c$
studied. We study 2VPF for $4096^2$, $2048^2$, $1024^2$, $512^2$,
$256^2$, $128^2$ and $64^2$ data points. All length scales which
do not satisfy the criteria $P_0 < 1/e$ were excluded from
analysis to avoid lattice effects. We also study different levels
of nonlinearity and cell sizes as in the analysis of VPF. Dashed
lines correspond to the theoretical prediction $\tau^2/ y^2
\propto y^{-(1 + \omega)}$, where $-\omega$ is also the slope of
one point VPF.}
\end{figure*}

\subsection{Simulations}

The simulations used here are numerical models for the
gravitational clustering of  collisionless  particles in an
expanding background. We study evolution of initial
two-dimensional Gaussian perturbations in $\Omega = 1$ universe.
All the 2D simulations are done with a particle-mesh (PM) code
with $4096^2$ particles with an equal number of grid points. The
code has at least twice the dynamical resolution of any other PM
code with which it has been compared. The 2D simulations we use
here are very similar to those described in detail in Beacom et.
al (1991),  a video of their evolution in shown in Kauffmann \&
Melott (1992), but with $64$ times more particles. In this paper
we analyze a subset of the simulations with featureless power-law
initial spectra of the general form,
\begin{eqnarray}
P(k)&\propto&\ k^n\ {\rm for}\ k \le k_c,\nonumber \\ &=&~ 0\ {\rm
for}\ k
> k_c.
\end{eqnarray}
We have analysed power-law models with power spectral index $n=
2,1, 0, -1$ in 2D, with a cutoff in each case at the Nyquist wave
number $k_c = 2048\,k_f$,  where $k_f=2 \pi/ L_{\rm box}$ is the
fundamental mode associated with the box size.

We choose $\sigma(k_{\rm NL})$, the epoch when the scale
$2\pi/k_{\rm NL}$ is going nonlinear, as a measure of time. The
first scale to go nonlinear is the one corresponding to the
Nyquist wave number. This happens, by definition, when the
variance $\sigma$ is unity. As $\sigma$ increases, successive
larger scales enter in the nonlinear regime. The simulations were
stopped at $\lambda_{\rm NL}=2l_{\rm grid}, \ 4l_{\rm grid},\
8l_{\rm grid}, ....,\ L_{\rm box}/2$. In our study  we have
studied epochs till $L_{\rm box}/16$ goes nonlinear.

The construction of $h(x)$ uses measurements of the CPDF from
several different epochs. We concentrated on smaller scales which
are already in the highly nonlinear regime. Cell sizes of
$(4l_{\rm grid})^2$, $(8l_{\rm grid})^2$ and $(16l_{\rm grid})^2$
were studied for four different epochs when $L_{\rm box}/8$,
$L_{\rm box}/16$, $L_{\rm box}/32$ and $L_{\rm box}/64$ have gone
nonlinear.

The growth rates of various modes in the linear regime were
studied by Melott et al. (1988) for this PM code. The results at
$\lambda = 3l_{\rm grid}$ are equivalent to those obtained by a
typical PM code at $\lambda = 8l_{\rm grid} $, owing to the
staggered mesh scheme. So we expect that our code performs well at
the wavelength associated with four cells and since the collapse
of $4\,l_{\rm grid}$-size perturbations will give rise to
condensations of diameter $ 2\,l_{\rm grid} $ or less, the
smallest cell size that can be safely resolved is $ 2\,l_{\rm
grid}$. The epochs we include in our study are not affected by the
fact that simulations were started by Zel'dovich approximation.

\subsection{Data Analysis}

For analysing the data, computations of the count probability
distribution function (CPDF) $P_l(N)$ and 2CPDF $P_{l,r}(N_1,N_2)$
were performed by laying down a grid of mesh spacing $l$ and
counting the occupation number in each cell; this yields the
probability of finding $N$ particles in cell size $l$ and also the
joint probability of finding $N_1$ and $N_2$ particles in two
different cells separated by a distance $r$. Statistics were
improved by perturbing the grid in each orthogonal direction and
keeping the mesh undistorted while repeating the counting process.
We considered cells of size $4l_{\rm grid}$, $8l_{\rm grid}$ and
$16l_{\rm grid}$ in our studies. With these cell sizes we can
reach probabilities as small as few times $10^{-9}$. We have also
checked that our results do not change if we increase the number
of sampling cells; we get exactly the same results even with ten
times  fewer cells. For computing the VPF we have only considered
cell sizes where the condition $P_0 > 1/e$ is satisfied; it was
shown by Colombi et al. (1995) that the properties of voids larger
than this  are affected by the grid which is used to start
simulations.

To extend the range of $N_c$ for which we can study the scaling
properties we have diluted the data while studying scaling
properties of VPF. Different levels of dilution such as $2048^2$,
$1024^2$, $512^2$ and $256^2$ were considered. Results of the VPF
analysis are plotted in Figure 5, showing that the scaling ansatz
predicts the VPF very well. We find a power-law profile for
$\sigma(N_c)$ for large values of $N_c$. At small values of $N_c$,
when the data is dominated by Poisson noise, $\sigma(N_c)$ tends
to unity. Our results of analysis of VPF proves beyond doubt that
the hierarchical ansatz is indeed a good approximation in the
highly non-linear regime. Since $\sigma(N_c)$ acts as a generating
function for the $S_N$ parameters this means that all $S_N$
parameters reach a constant value in this limit. Furthermore, as
the $S_N$ parameters are linear combinations of product of powers
of amplitudes of different vertices that appear in the tree-level
correlation hierarchy, this will also mean that these vertices too
are constant in highly non-linear regime.

However, we have also found a significant departure from such
scaling for $n=-2$ spectra. The maximum value of $N_c$ for which
the VPF can be probed reliably decreases with increasing power on
larger scales and the simulation grid, which remains undistorted
in underdense regions for spectra with more power at larger scales
even in the highly non-linear regime. these two factors may be
responsible for a spurious departure from scaling in this case.
Strong motivation for suspecting that the apparent departure  we
detect in scaling for VPF for $n=-2$ may not be real is that we do
not detect any  corresponding effect in  $h(x)$.

We use four neighbouring equidistant cells in a grid to evaluate
the bias with respect to the central cell. We adopt two different
methods to evaluate the bias  which, as suggested by Bernardeau
(1994), test the factorizability property of bias of over-dense
cells predicted by the hierarchical ansatz. In the first method we
compute the bias associated with each overdense cells with respect
to background mass. This is done by casting the definition of bias
in a following way:
\begin{equation}
b(>N) = \Big ( { \sum_{N_1>N} P(N_1)b(N_1) \over \sum_{N_1 > N}
P(N_1)}  \Big ) =  { 1 \over \xi_{ij}}\Big ( {{{1 \over N_2}
\sum_{N_1 > N} \sum_{N_2} N_2 P(N_1, N_2)} \over \sum_{N_1 > N}
P(N_1)} - 1 \Big ).
\end{equation}
In the second method we directly compute the bias from correlation
of two  different over-dense cells:
\begin{equation} b(>N) = \Big ( {
\sum_{N_1>N} P(N_1)b(N_1) \over \sum_{N_1 > N} P(N_1)}  \Big ) =
\sqrt{ \left({ 1 \over \xi_{ij}}\Big ( {{ \sum_{N_1 > N} \sum_{N_2
> N}  P(N_1, N_2)} \over (\sum_{N_1 > N} P(N_1))^2} - 1 \Big )
\right)}
\end{equation}
In both cases we compute the bias for cells whose occupancy is
greater than some particular threshold $x$ and then we vary the
threshold to study $b(>x)$ for different values of $x$. We
initially attempted to compute directly the function $b(x)$ but we
found measurements of the cumulative distribution $b(>x)$ to be
much more stable. We also found that the bias computed from both
methods mentioned above matches, and also the slope of the bias
function for  overdense cells agrees well with theoretical
predictions. In particular, the bias increases with $x$ as a power
law with the index of power-law being less than unity, whereas for
highly overdense cells the bias is directly proportional to $x$.
Determination of the bias is more difficult for spectra with more
power on smaller scales, due to absence of long range correlations
in such models.

Computation of the 2VPF was done using a similar technique as in
the computation of bias. We used four neighbouring cells
for every cell to find the joint probability of finding the pair
of cells to be devoid of any particles; we did this for several
different separations of two cells. For a given value of the
scaling parameter $N_c$, we have also studied the scaling
properties associated with the 2VPF for larger separation. Data
was diluted to  from $4096^2$ to $2048^2$, $1024^2$, $512^2$ and
$256^2$ to increase the available dynamic range of the scaling
parameter $N_c$ as we did for the case of VPF. For large cell
separations, where all higher order terms in $\xi_{12}/\bar \xi$
are negligible,
\begin{equation}
{1\over {\xi_{12}}}{1 \over {\bar N}^2}\rm \log \left({P(0,0)
\over P(0)^2} \right)= {\tau^2(N_c) \over N_c^2}.
\end{equation}
We found that the scaling function $\tau(N_c)^2/N_c^2$, which
encodes the scaling properties of the 2VPF and also acts as a
generating function for cumulant correlator (Munshi \& Melott
1998) is described very accurately by the theoretical prediction,
made on the basis of the hierarchical ansatz; see Figures 6 \& 7.

\begin{table}
\begin{center}
\caption{Parameters of the fitting function $h(x)$ and $b(x)$ in 2D}
\label{tabhx3D}
\begin{tabular}{@{}lccccc}
& -1& 0 & 1 & 2\\
$\omega$& .35 & .67 & .85& .9\\
a&1.25 & 1.778 & 2.81& 4.46
\end{tabular}
\end{center}
\end{table}

\section{Finite Volume Correction}
The finite size of galaxy catalogues and numerical simulations
always affects the determination of many-body correlation
functions from real or simulated samples. The higher the order of
correlation measured, the greater the contribution to it from
exceptionally dense and consequently rare clumps.  Their
determination is therefore always affected by the presence or
absence of such clumps in a finite sample. This manifests itself
in a sharp cutoff in all the one-point functions we have defined
such as $P(N)$, $b(N)$, and $\nu_2(N)$. Using the scaling
arguments we have already described n it is however
 possible to supplement the measured quantities with the functions
by $h(x)$, $b(x)$, and $\nu_2(x)$. These quantities have been
computed by series expansion of multi-point VPF in increasing
power of $(\xi_{ij}/ \bar \xi_i)$ for second and third order
by Bernardeau \& Schaeffer (1992) and for fourth and fifth order
in our Paper I. Using their results we show here that scaling
arguments once verified at lower order in $(\xi_{ij}/\bar \xi_i)$
can be used to estimate finite volume errors for in estimates of
the multi-point cumulant correlators.

If $N_{\rm max}$ is the maximum occupancy of cells of certain
length scales in a catalog, measured $P(N)$ for $N<N_{\rm max}$
shows an abrupt cutoff for $N=N_{\rm max}$. We supplement the
measured $P(N)$ by $\Pi(\nu)$ for $N>N_{\rm max}$ to construct the
corrected $P^{\rm c}(N)$ (Munshi et al. 1998c); $\Pi(\nu)$ can be
expressed in terms of the scaling function $h(x)$ describing
$P(N)$:
\begin{eqnarray}
P^{\rm c}(N) & = & \Big ( 1 - {{\cal A} \over \bar \xi_2} - {\cal
B} { (N - \bar N) \over N_c} \Big )P(N) ~~~~~~~~N < N_{\rm max} \\
\nonumber P^{\rm c}(N) &  = &
\Pi(\nu)~~~~~~~~~~~~~~~~~~~~~~~~~~~~~~~~~~~~~~~~N > N_{\rm max}.
\end{eqnarray}
The corrected CPDF satisfies two constraint equations similar to
the case of the uncorrected CPDF. The constraints
\begin{equation}
\sum P(N) =1, ~~ \sum P^{\rm c}(N) =1,~~ \sum NP(N) = \bar N,~~ \sum
NP^{\rm c}(N) = \bar N
\end{equation}
can be used to determine the constants ${\cal A}$ and ${\cal B}$
where
\begin{eqnarray}
{\cal A} & = & H_0 = \int h(x)  dx  \\ {\cal B} & = & H_1 = \int x
h(x) dx
\end{eqnarray}
and
\begin{equation}
{\Sigma_p}^{\rm c} = \Sigma_p - (p+1) H_1 \Sigma_{p+1}^{\nu_s} +
{H_p \over p!}.
\end{equation}

Similarly, the bias associated with overdense cells $b(N)$ shows
an abrupt cutoff at $N>N_{\rm max}$ which we supplement by its
continuous analogue $b(\nu)$, which can be expressed in terms of
its scaling function $b(x)$:
\begin{eqnarray}
P^{\rm c}(N)b^{\rm c}(N) & = &  \Big ( 1 - {\cal C}- {\cal D} \big
(N - {{\langle N^2 \rangle}_b\over \bar N} \big ) \Big )P(N)b(N)
~~~~~~~N < N_{\rm max} \\ \nonumber P^{\rm c}(N)b^{\rm
c}(N) & = &  \Pi(\nu) b(\nu)
~~~~~~~~~~~~~~~~~~~~~~~~~~~~~~~~~~~~~~~~~~~~ N > N_{\rm max},
\end{eqnarray}
where we have denoted $\sum N^2P(N)b(N)$ by $\langle N^2
\rangle_b$. The unknown parameters ${\cal C}$ and ${\cal D}$ can
be determined by using the fact that both $b(N)P(N)$ and corrected
$b^{\rm c}(N)P^{\rm c}(N)$ satisfy similar constraints. For the
zeroth moment: $\sum b(N)P(N) = 0$, $\sum b^{\rm c}(N)P^{\rm
corr}(N) = 0$ and similarly for the first moment we have $\sum N
b(N)P(N) = \bar N$, $\sum N b^{\rm c}(N)P^{\rm c}(N) = \bar N$.
The constraint on the zeroth-order moment can be used to determine
${\cal D}$, and the first-order constraint fixes the value of
${\cal C}$:
\begin{eqnarray}
{\cal C}&   = & B_1 = \int_{N_{\rm max}/N_c}^{\infty} x h(x) b(x)
dx
\\ {\mathcal D} &  = & B_0 = \int_{N_{\rm max}/N_c}^{\infty} h(x) b(x) dx
\end{eqnarray}
and
\begin{equation}
{\Sigma^{b}_p}^{\rm c} = ( 1 - B_1)\Sigma^b_p - (p+1) B_0
{\Sigma^b_{p+1}} + {B_p \over p!}.
\end{equation}
Where we have defined the $p^{th}$ order moment of $b(x)h(x)$ by
$B_p$ and
\begin{equation}
B_p = \int_{N_{\rm max}/N_c}^{\infty} x^p b(x)h(x) dx.
\end{equation}

Extending these calculations to the case of MCPDFs is fairly
straightforward. The constraints  $\sum P(N)\nu_s(N) = \bar N
\nu_s$, $\sum NP^{\rm c}(N)\nu_s^{\rm c}(N) = \bar N \nu_s$ and
$\sum P(N)\nu_s(N) =1$, $\sum P^{\rm c}(N)\nu_s^{\rm c}(N) =1$ can
be used to determine the values of ${\cal X}$ and ${\cal Y}$
appearing in the renormalization of $P(N)\nu_s(N)$ that corrects
for the finite volume effect:
\begin{eqnarray}
P^{\rm c}(N)\nu_n^{\rm c}(N) & = & \Big ( 1 - {\cal X} -
{\cal Y} \big (N - {{\langle N^2 \rangle}_{\nu_n}\over \bar N}
\big ) \Big )P(N)\nu_n(N)  ~~~~~~~~N < N_{\rm max} \\ \nonumber
P^{\rm c}(N)\nu_n^{\rm  c}(N) & = & \Pi(\nu) \nu_n(\nu)
~~~~~~~~~~~~~~~~~~~~~~~~~~~~~~~~~~~~~~~~~~~~~~ N > N_{\rm max},
\end{eqnarray}
where ${\cal Y}$ and ${\cal X}$ are the first and zeroth order
moments of $h(x)\nu_n(x)$ respectively and $\langle N^2
\rangle_{\nu_n}$ represents $\sum N^2 P(N)\nu_n(N)$; these
quantities can be expressed by
\begin{eqnarray}
{\cal X} & = & Z_1^{\nu_n} = { 1 \over
\nu_n}\int_{N_{max}/N_c}^{\infty} x h(x) \nu_n(x) dx  \\ {\cal Y}
= Z_0^{\nu_n} & = & {1 \over \nu_n}\int_{N_{\rm max}/N_c}^{\infty}
h(x) \nu_n(x) dx
\end{eqnarray}
and
\begin{equation}
{\Sigma^{\nu_n}_p}^{c} = (1 - Z_1^{\nu_n})\Sigma_p^{\nu_n} -
Z_0^{\nu_n}(p+1) \Sigma_{p+1}^{\nu_n} + {Z_p \over p!},
\end{equation}
where we have denoted the $p^{th}$ moment of $\nu_n(x)h(x)$ by
\begin{equation}
Z_p^{\nu_n} = {1 \over \nu_n}\int_{N_{max}/N_c}^{\infty} x^p  h(x)
\nu_n(x) dx.
\end{equation}

If we also notice that  we  have to correct $N_c$ for the finite
volume effect using $N_c^{\rm c} = N_c( 1 - 6H_1 \Sigma_3 + H_2)$.
Incorporating this correction we can finally write
\begin{eqnarray}
{\Sigma_p}^{\rm c} & = & {\Sigma_p - (p+1) H_1
\Sigma_{p+1}^{\nu_s} + {H_p \over p!} \over ( 1 - 6H_1 \Sigma_3 +
H_2)^{p-1}} \\ {\Sigma^{b}_p}^{\rm c} & = & {( 1 - B_1)\Sigma^b_p
- (p+1) B_0 {\Sigma^b_{p+1}} + {B_p \over p!} \over ( 1 - 6H_1
\Sigma_3 + H_2)^{p-1}} \\ {\Sigma^{\nu_n}_p}^{\rm c} & = & {(1 -
Z_1^{\nu_s})\Sigma_p^{\nu_n} - Z_0^{\nu_n}(p+1)
\Sigma_{p+1}^{\nu_n} + {Z_p^{\nu_s} \over p!} \over ( 1 - 6H_1
\Sigma_3 + H_2)^{p-1}}.
\end{eqnarray}
We have kept only terms which are leading order in $1/\bar \xi$ in
the expressions and hence they are valid only in the highly
nonlinear regime. We have also ignored a term in  $\langle N^2
\rangle_{\nu_s}/\bar \xi_2 \nu_n \bar N^2$ since we expect
$\langle N^2 \rangle_{\nu_n}/\nu_n \bar N^2$ to be of order unity.
The inclusion of terms representing loops of lower-order diagrams
in the calculations will bring not only moments of $b(x)h(x)$ and
$\nu_n(x)h(x)$ but also moments of powers of these scaling
functions which we have neglected here.

Cumulants and cumulant correlators have been measured with
presently available N-body and galaxy catalogues. Although several
analysis for finite volume correction for cumulants are available
in the literature (Colombi et al. 1992, 1994, 1995, 1996, Munshi et al.
1998), no such analysis has been
done for cumulant correlators. Our analysis based on the
hierarchical ansatz provides an estimate for finite volume
correction and also can be used to correct such effects.
Implementation of our method for measurements of MCCs will be
presented elsewhere.

It was shown by Szapudi \& Szalay (1997) that cumulant correlators
can actually be used to separate amplitudes associated with
different tree topologies at low order. However, at higher orders
the number of equations becomes less than the number of
independent tree topologies rendering the system indeterminate.
This can only be cured if we increase the number of points by
using multi-point cumulant correlators. This in principle will
allow us not only to determine the amplitudes associated with tree
topologies of arbitrary order but also allow an additional
consistency check because the  number of variables is less than
the number of equations they have to satisfy. Of course going from
two-point cumulant correlators to multi-point cumulant correlators
makes numerical computation more complicated, especially because of
the shape dependence of multi-point cumulant correlators for more
than two points. With the availability of simulations with much
larger dynamic range a determination of the $\nu_n$ parameters
will become possible in the reasonably near future.

The procedure which we have introduced here for finite volume
corrections however requires {\em a priori} knowledge of the
$\nu_n$ parameters which one needs to determine from the same
sample. This calls for an iterative procedure which can be applied
until the whole process becomes convergent.

\section{Conclusions}
In this and the previous paper (Paper I), we have investigated
properties of the multi-point cumulant correlators - the natural
generalisations of the one-point cumulants and the two-point
cumulant correlators. We also developed a method which can be used
to determine the amplitudes associated with tree diagrams of
different topologies, without making any simplistic assumptions
about them. This method can be used to test any models of
gravitational clustering in the highly non-linear regime, such as
the hierarchical scaling ansatz.

We have developed a diagrammatic method to represent the
decomposition of multi-point cumulant correlators in terms of
factorial moments and factorial correlators. Using this approach,
we have found that every higher-order contribution to
$(\xi_{ij}/\bar \xi_i)$ introduces a new function which
corresponds to a new vertex in the tree-level hierarchy, labeled
by $\nu_n(N)$. It was shown in Paper I that these functions
display similar scaling relations as the functions $P(N)$ and
$b(N)$. We have used these scaling relations for $\nu_n(N)$ to
estimate the effect of the finite size of N-body simulations or
galaxy catalogues in estimates of statistical descriptors, such as
cumulants, cumulant correlators and multi-point cumulant
correlators.

We have tested various predictions of the hierarchical ansatz,
relating to one-point and two-point quantities in the highly
non-linear regime, using high-resolution 2D numerical simulations.
We  find that the scaling functions $h(x)$ and $\sigma(N_c)$ can
describe the behaviour of the CPDF and VPF in highly non-linear
regime for all length scales, as the hierarchical assumption
predicts. These functions were, however, found to depend on
initial conditions even in the highly nonlinear regime. As
expected from general scaling arguments, it was found that
$\sigma(N_c)$ attains a asymptotic power-law $N_c^{-\omega}$ for
large values of $N_c$. However, such models are not able to
predict the values of $\omega$; more detailed dynamical arguments
would have to be developed to make such a prediction. Numerically,
however,  we have found that $\omega$ depends on the form of the
initial power spectrum, increasing with the relative amount of
small-scale power. The index $\omega$ also enters in the
description of $x^2h(x)$ which we found to have a power law
profile with power law index $\omega$ and an exponential cutoff
for large values $x$. This transition occurs around the typical
cell occupancy in overdense cells where $N=N_c$, i.e. $x=1$.

Extending such studies of one-point quantities for the first time
to their two-point analogues such as 2CPDF and 2VPF, we were able
to check how overdense cells are biased with respect to the
underlying mass distribution. We find, in accordance with
 hierarchical predictions, that the bias is an intrinsic property
of  collapsed objects.  One can associate scaling variable $x$
with every collapsed object, which is a function of cell occupancy
$N$, radius of the cell, and the  variance of matter distribution
$\bar{\xi}_2$ on that length scale. We also found that the bias is
factorizable and for two different objects with different values
of scaling parameters $x_1$ and $x_2$ and with intrinsic bias
$b(x_1)$ and $b(x_2)$, it is possible to express their bias with
respect to each other as $b(x_1)b(x_2)$.

Elsewhere in the literature,  extensions of the Press-Schechter
theory have been used to compute the bias of collapsed objects (Mo
\& White 1996) and also their higher-order one-point moments such
as $S_N$ parameters (Mo et al. 1997). These are in agreement with
the results we have obtained, especially those presented in Paper
I. However, it remains unclear whether the generalisation of this
formalism to two-point quantities (such as the CCs) will produce
accurate predictions. Other works which have focussed on
predicting bias and higher-order correlation functions are related
to extensions of the Zel'dovich approximation (Lee \& Shandarin
1997,1998, Catelan et al. 1997). It would be a very important test
of the usefulness of these, and other approximation schemes for
non-linear gravitational evolution to see if they can actually
reproduce the multi-point statistics of collapsed objects as well
as the simple hierarchical ansatz we have discussed here.

 All these theoretical predictions were made
in the lowest order of $(\xi_{ij}/ \bar \xi_i)$, but were found to
describe simulation results even when two cells are not separated
by large distances.

Just as the VPF is related to the generating function of the $S_N$
parameters, the 2VPF is related to the generating function of the
cumulant correlators (CCs). In earlier numerical studies
(Munshi \& Melott 1998)  we have shown that, in general, CCs are
factorizable, so that $C_{pq} = C_{p1}C_{q1}$, when two cells are
separated by a large distance. Such a decomposition of CCs is
related to the  factorization of the 2VPF which we have
established in this paper, thus confirming our earlier findings.
The factorization property of bias is also a direct consequence of
the scaling properties we study here, so our results on the
subject of  bias also vindicate our earlier findings.

Our numerical study was done in 2D and we plan to extend such
analysis in case of 3D in near future, incorporating studies of
higher-order correlations of over-dense cells.

It is interesting to note that although we have demonstrated that
the hierarchical ansatz (62) is a very good approximation to the
behaviour of higher-order correlation functions, stable clustering
is known to be violated in 2D in highly non-linear regime
(Munshi et al. 1998a). It is often argued that stable
clustering is a necessary ingredient in the hierarchical ansatz.
This appears not to be the case.  We shall discuss this issue in
more detail in a future paper.

Finite sample (or simulation) size known to introduce a sharp
cut-off in the CPDF for $N = N_{\rm max}$ corresponding to the
density of the densest densest cell in the catalog. Extending
methods developed earlier (Munshi et al. 1998c) for correcting
finite volume effects in the CPDF $P_l(N)$ by supplementing the
missing information beyond $N=N_{\rm max}$ from scaling function
$h(x)$ (which can then be used to compute the $S_N$ parameters),
we developed a similar method for MCPDF in this paper. At the
level of 2CPDF e.g., one can use the scaling property of $b(N)$
which is encoded in scaling function $b(x)$ to correct or estimate
finite volume effects in the extraction  of CCs. We show that this
technique can be generalised in a straightforward manner to the
case of MCPDFs. At present there is no other general formalism to
estimate the finite volume error for multi-point statistics. Our
method provides the first step in this direction.

\section*{Acknowledgments}
Dipak Munshi acknowledges support from PPARC under the QMW
Astronomy Rolling Grant GR/K94133. Peter Coles received a PPARC
Advanced Research Fellowship during the period when much of this
work was completed. We are grateful for support under the
NSF-EPSCoR program, as well as the Visiting Professorship and
General Research Funds of the University of Kansas.

\end{document}